\documentclass[%
reprint,
superscriptaddress,
a4paper,
showpacs,preprintnumbers,
amsmath,amssymb,
aps,
prstab,
floatfix,
]{revtex4-1}

\usepackage{graphicx}
\usepackage[caption=false]{subfig}
\usepackage{hyperref}
\hypersetup{colorlinks=true,citecolor=blue}

\begin{document}
\preprint{APS/XX-XXX}
\title{2D-Oide effect}
\thanks{This work has been financially supported by The French National Center for Scientific Research (CNRS) and The European Organization for Nuclear Research (CERN).}%
\author{O.R. Blanco}
\affiliation{LAL, Universit\'e Paris-Sud, CNRS/IN2P3, Orsay 91898, France}
\affiliation{CERN, Geneva 1211, Switzerland}
\author{R. Tom\'as}
\email{rogelio.tomas@cern.ch}
\affiliation{CERN, Geneva 1211, Switzerland}
\author{P. Bambade}
\affiliation{LAL, Universit\'e Paris-Sud, CNRS/IN2P3, Orsay 91898, France}
\date{\today}

\begin{abstract}
The Oide effect considers the synchrotron radiation in the final focusing quadrupole and it sets a lower limit on the vertical beam size at the Interaction Point, particularly relevant for high energy linear colliders. The theory of the Oide effect was derived considering only the radiation in the focusing plane of the magnet.\par
This article addresses the theoretical calculation of the radiation effect on the beam size considering both focusing and defocusing planes of the quadrupole, refered to as 2D-Oide. The CLIC~3~TeV final quadrupole (QD0) and beam parameters are used to compare the theoretical results from the Oide effect and the 2D-Oide effect with particle tracking in PLACET. The~2D-oide demonstrates to be important as it increases by 17\% the contribution to the beam size.\par
Further insight into the aberrations induced by the synchrotron radiation opens the possibility to partially correct the 2D-Oide effect with octupole magnets. A beam size reduction of 4\% is achieved in the simplest configuration, using a single octupole.\par
\end{abstract}
\pacs{29.20.Ej, 41.85.Lc}
\maketitle
\section{Introduction}\label{Oideeffect}
\begin{figure}[h]
\includegraphics[width=0.48\textwidth]{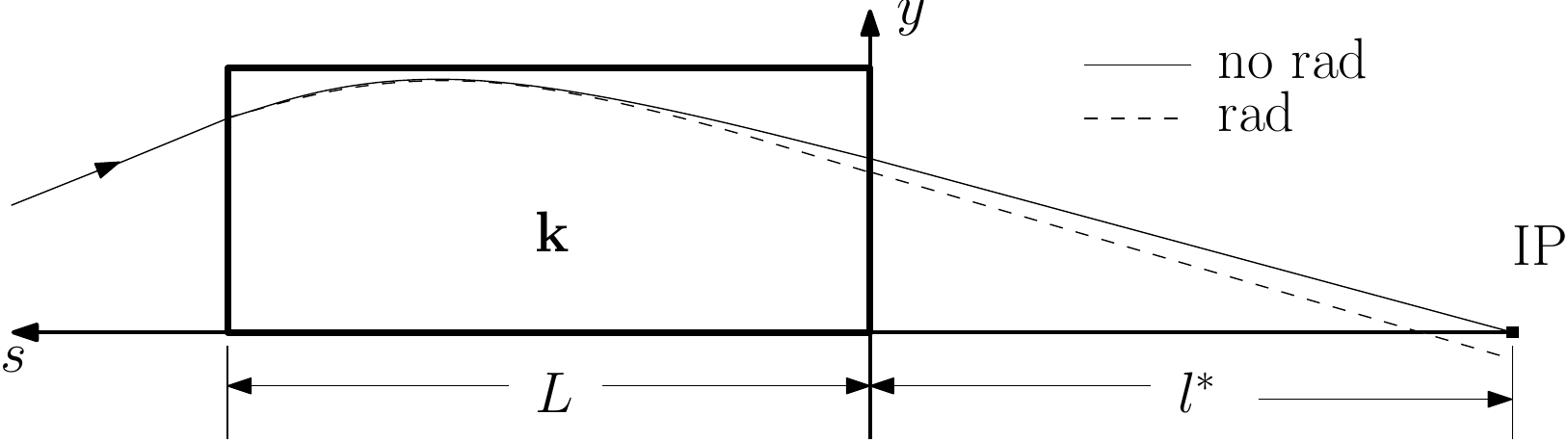}\caption{The solid and dashed lines represent the particle trajectory without and with radiation, respectively.}\label{f:Oideeffect}
\end{figure}

Synchrotron radiation in a focusing quadrupole magnet of length~$L$ and gradient~$k$, schematically represented in Fig.~\ref{f:Oideeffect}, changes the energy of the particle and modifies the focusing effect. This results in a limit on the minimum beam size at the Interaction Point~(IP) located at a distance~$l^*$ from the quadrupole. This is referred to as Oide effect~\cite{Oide}.\par
 The beam size growth due to radiation is added quadratically to the linear beam size $\sigma_0^2=\epsilon\beta^*$ where $\beta^*$ represents the optical beta function at the IP and $\epsilon$ is the emittance. Therefore, $ \sigma^2 = \sigma_0^2 + \sigma_{\text{oide}}^2$, where the beam size contribution from the Oide effect is \cite{Oide},
 \begin{equation}
  \sigma^2_{\text{oide}} = \frac{110}{3\sqrt{6\pi}}r_e\frac{\lambda_e}{2\pi}\gamma^5 F(\sqrt{k}L,\sqrt{k}l^*)\left(\frac{\epsilon}{\beta^*}\right)^{5/2},
  \label{Oideequ}
 \end{equation}
 with
 \begin{align}
  F\left(\sqrt{k}L, \sqrt{k}l^*\right) = &\int_0^{\sqrt{k}L}|\sin\phi+\sqrt{k}l^*\cos\phi|^3\notag\\
  &\left[\int_0^\phi(\sin\phi'+\sqrt{k}l^*\cos\phi')^2 d\phi'\right]^2d\phi\label{OideF}
 \end{align}
and $\lambda_e$ is the Compton wavelength of the electron, $r_e$ is the classical electron radius and $\gamma$ is the relativistic factor.\par
The primitive of the double integral used to calculate $F$ in Eq.~(\ref{OideF}) is derived analytically in \cite{Blanco:1967497} with the goal to increase the computational calculation speed. It has been included in MAPCLASS2~\cite{Mapclassorig,Tomas:944769,Martinez:1491228,githubMapClass2} to be used in lattice design and optimization.\par
Although the total contribution to beam size depends on the lattice and beam parameters, the minimum achievable beam size is given by \cite{Oide}
\begin{equation}
 \sigma_{\text{ min}} = \left(\frac{7}{5}\right)^\frac{1}{2}\left[\frac{275}{3\sqrt{6\pi}}r_e\frac{\lambda_e}{2\pi}F\left(\sqrt{K}L,\sqrt{K}l^*\right)\right]^\frac{1}{7}(\epsilon_{N})^\frac{5}{7},
\end{equation}
where $\epsilon_N=\gamma\epsilon$ is the normalized emittance, showing the independence from beam energy.\par
{\renewcommand{\arraystretch}{1.2}
\begin{table*}[thp]
\begin{tabular}{lcccccccccc}\hline\hline
Lattice &$\epsilon_N$& $\gamma$& $\sigma_0$&$k$&$L$&$l^*$& $F$ & $\sigma_{\text{oide}}$&$\sigma$&$\sigma_{\text{min}}$\\
 &[nm]&[$10^3$]&[nm]&[m$^{-2}$]&[m]&[m]&&[nm]&[nm]&[nm]\\\hline
CLIC 3 TeV & 20 & 2935.0 & 0.70 & 0.116 & 2.73 &3.5&  4.086  & 0.87 & 1.10& 1.00 \\
CLIC 500 GeV & 25 & $\;\;$489.2 & 2.3\;\; & 0.077 & 3.35 &4.3& 4.115 & 0.08 & 2.3 & 1.17\\
ILC  500 GeV & 40 & $\;\;$489.2 & 5.7\;\; & 0.170 & 2.20 &4.3& 9.567 & 0.04 & 5.7 & 1.85\\\hline\hline
\end{tabular}\caption{Vertical beam size and radiation beam size contribution for three lattices. $\epsilon_N$ is the normalized emittance, $\epsilon_N=\gamma\epsilon$.}\label{t:Sigmas}
\end{table*}
}
Table~\ref{t:Sigmas} shows relevant parameters of the last vertically focusing magnet QD0 ($k$, $L$), the beam and lattice optics ($\epsilon_N$, $\gamma$, $\sigma_0$, $l^*$) for the two main linear collider projects ILC~500~GeV~\cite{ILCdes}, and CLIC at 500~GeV and 3~TeV~\cite{CLICdes}. These are used to show 
that the contribution of the Oide effect to vertical beam size is only significant for CLIC~3~TeV. Also, columns $\sigma$ and $\sigma_\text{min}$, show that the final vertical beam size is comparable to the minimum achievable.\par
\par
If none of the beam parameters or $l^*$ is to be changed then $F$ can be used as a figure of merit of the quadrupole set-up as it is calculated only from $k$, $L$ and $l^*$, where the target is to reduce $F$ as much as possible. The standard procedure of reducing $F$ is by increasing the length of the quadrupole and reducing its gradient. In the following this is illustrated for CLIC~3~TeV.\par
\begin{figure}[floatfix]
\subfloat[]{\label{fig-3TeV:a}%
\includegraphics[width=0.33\textwidth,angle=-90]{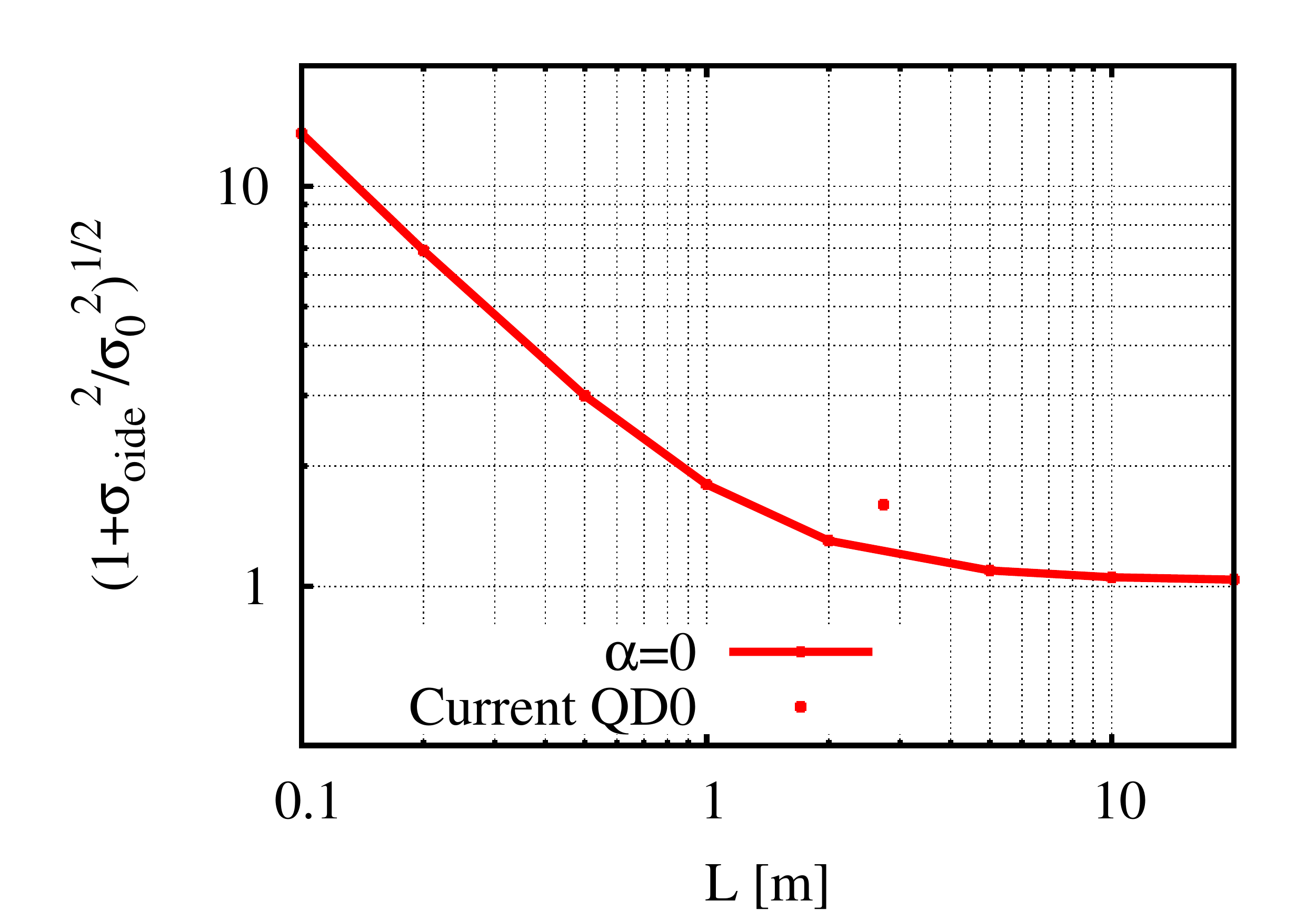}
}\\
\subfloat[]{\label{fig-3TeV:b}%
\includegraphics[width=0.33\textwidth,angle=-90]{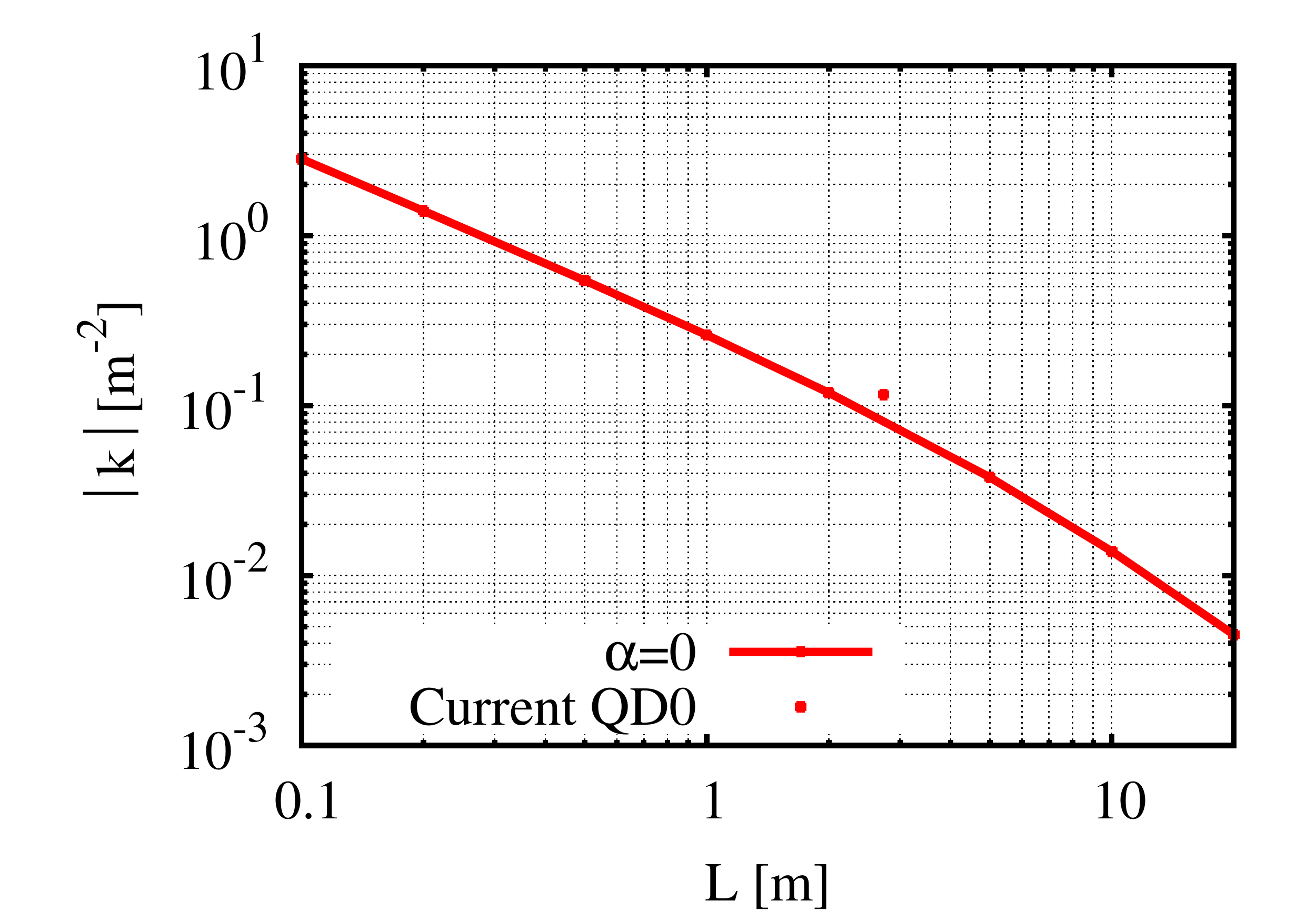}
}
\caption{Oide effect beam size contribution for CLIC~3~TeV design parameters. (a) Total beam size normalized to designed linear beam size as a function of quad length for the minimum focusing $k$ and the current QD0. (b)~$k$~in~the~two previous cases.}\label{fig-3TeV}
\end{figure}

The relative increase in the beam size due to the Oide effect is given by 
\begin{equation}
\sigma/\sigma_0=\sqrt{1+\sigma_{\text{oide}}^2/\sigma_0^2}.\label{eq:oidesq}
\end{equation}
Figure~\ref{fig-3TeV}(a) shows the the relative increase due to Oide effect 
when $k$ is set to cancel the optics function alpha, $\alpha$, just at the quadrupole opposite face to the IP, the particle input. This is done to set an absolute minimum in the Oide effect out of reach for realistic Final Doublet (FD) designs. Figure~\ref{fig-3TeV}(b) shows the $k$ values previously mentioned.\par
The current QD0 almost doubles the vertical beam size. It might be possible to reduce $\sigma^2_{\text{oide}}$ by a small factor by increasing the current quadrupole length and possibly using a lower $k$, however, this solution also points to increasing the lattice length. 
This could also generate difficulties in lattice design because of chromaticity and magnet stabilization. Quad lengths larger than 10~m do not lead to further significant improvements.\par
\section{2D-Oide effect}\label{Oide2D}
In order to gain understanding for the relevant case of CLIC at 3~TeV, Section~\ref{Oide2D} is dedicated to the theoretical derivation of the effect of radiation including the horizontal beam size and optics lattice parameters \mbox{(2D-Oide)}, and comparing results with particle tracking. This leads to a possible way to mitigate the effect, alternative to enlarging QD0, consisting in removing the correlation at the IP via normal octupole magnets. This is tested for the simplest case of one octupole magnet as an example.
\subsection{Analytical derivation}
Being $E_0$~the nominal energy of the beam and~$u$ the energy of the photon radiated, in a first order approximation where $u\ll E_0$, the total effect of radiation in the vertical displacement at the IP is calculated integrating along the magnet length as
\begin{equation}
\Delta y = \int^L_0\int_0^s f_y^2(\sqrt{k}s_1)\frac{u}{E_0}y'^*_0 ds_1 ds,\label{eq:deltays}
\end{equation}
where the inner integral represents the kick given by the radiated photon at a certain location $s$ propagated to the IP assuming a strong focusing in the vertical plane ($l^*~\gg~\beta^*_y$)~\cite{Oide}. $y'^*_0$ is the vertical particle angle at the IP without radiation and 
\begin{equation}
f_y(\phi)=\sin\phi+\sqrt{k}l^*\cos\phi.\label{eq:fy}
\end{equation}
Because $u$ is much smaller than $E_0$ in Eq.~(\ref{eq:deltays}), it is still possible to use the result in~\cite{Sands2} for the average energy loss per unit length in the magnet~$\bar{u}$ and its second moment~$\overline{u^2}$ shown in Eqs.~(\ref{eq:averu})~and~(\ref{eq:averu2}):
\begin{align}
\frac{\bar{u}}{E_0}=&\frac{2}{3}r_e\frac{\gamma^3}{\rho^2}\label{eq:averu} \\
\frac{\overline{u^2}}{E_0}=&\frac{55}{24\sqrt{3}}r_e\frac{\lambda_e}{2\pi}\gamma^5\frac{1}{|\rho|^3}\label{eq:averu2} 
\end{align}
This allows to calculate two important values: the mean effect of the radiation in the trajectory of one particle along the magnet $\overline{\Delta y}$ and its second moment $\overline{(\Delta y)^2}$.\par
\begin{figure}[floatfix]
\includegraphics[width=0.48\textwidth]{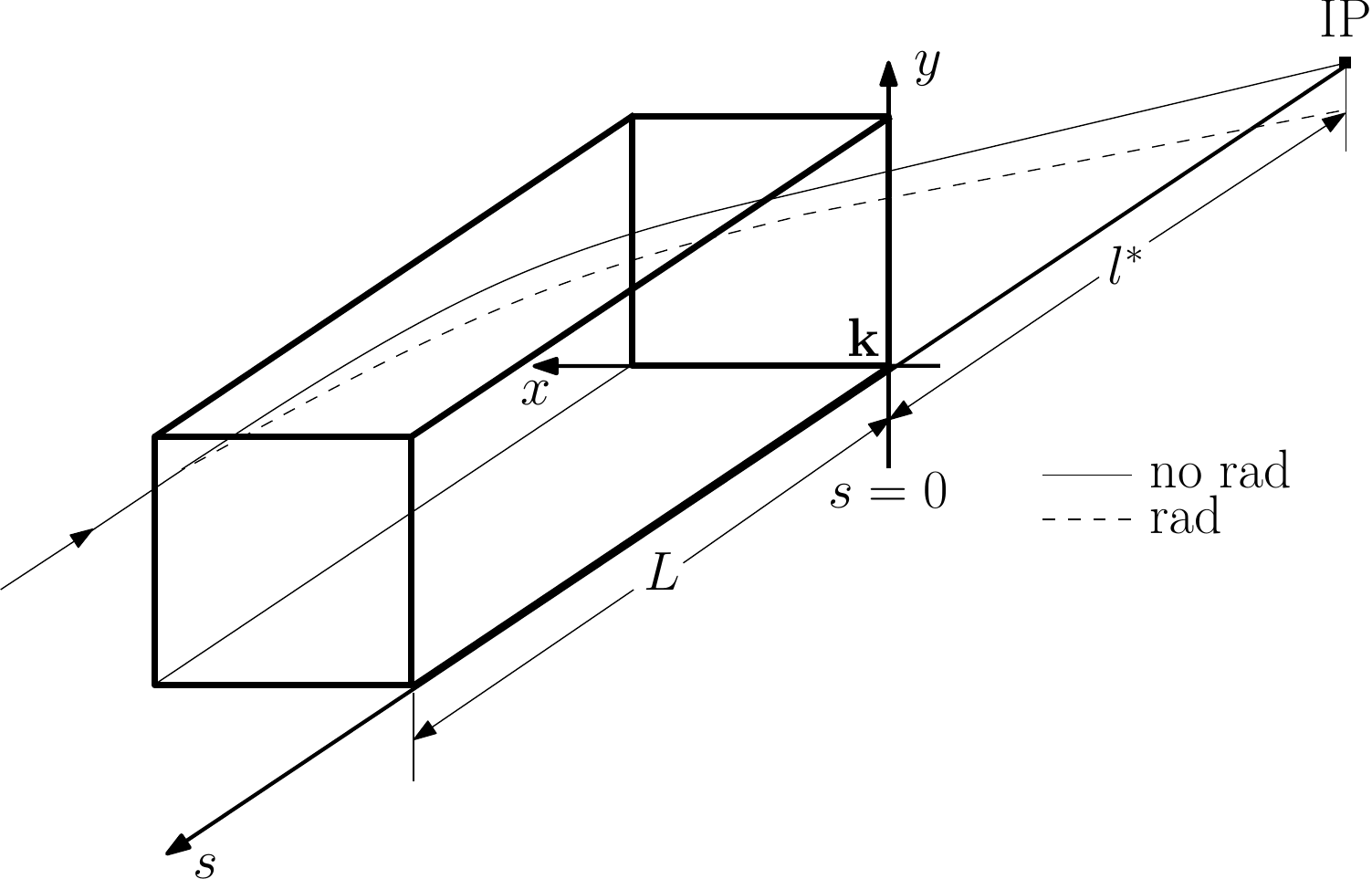}\caption{The solid and dashed lines represent the particle trajectory without and with radiation, respectively.}\label{f:Oide2D}
\end{figure}

The calculation derived by Oide considers the beam size in the defocusing plane of the final quadrupole to be negligible, however, if the beam size in the defocusing plane is considered, the bending radius of curvature~$\rho$ in the lens is given by
\begin{equation}
\frac{1}{|\rho(s)|}=\left|k\right|\sqrt{x^2(s)+y^2(s)}\label{eq:bendingr}
\end{equation}
and Fig.~\ref{f:Oide2D} shows schematically the particle trajectory. As shown in Eqs.~(\ref{eq:averu}) and (\ref{eq:averu2}), the radius of curvature depending on both transversal planes has a direct impact on the radiation.\par
\par
In addition, assuming $l^*\gg\beta^*_x$, the coordinates $x(s)$ and $y(s)$ in Eq.~(\ref{eq:bendingr}) can be calculated from the horizontal~$(x'^*_0)$ and vertical~$(y'^*_0)$ angles  at the IP as
\begin{align}
x(s) =& -x'^*_0 \frac{f_x(\sqrt{k}s)}{\sqrt{k}}\label{eq:greenx}\\
y(s) =& -y'^*_0 \frac{f_y(\sqrt{k}s)}{\sqrt{k}}\label{eq:greeny}
\end{align}
where
\begin{equation}
f_x(\phi)=\sinh\phi+\sqrt{k}l^*\cosh\phi\label{eq:fx}
\end{equation}
Combining the expressions in the Eqs.~(\ref{eq:averu}),~(\ref{eq:bendingr}),~(\ref{eq:greenx}) and (\ref{eq:greeny}), it is possible to average over the photon energy in Eq.~(\ref{eq:deltays}), obtaining:
\begin{equation}
\overline{\Delta y} = a (y'^*_0)^3+b(x'^*_0)^2y'^*_0\label{eq:deltay}
\end{equation}
where
\begin{align}
a =& \frac{2}{3}r_e\gamma^3\int_0^{\sqrt{k}L}f^2_y(\phi)F_y(\phi)d\phi=\frac{1}{3}r_e\gamma^3F^2_y\big(\sqrt{k}L\big)\label{eq:deltayintsa}\\
b=&\frac{2}{3}r_e\gamma^3\int_0^{\sqrt{k}L} f^2_x(\phi)F_y(\phi)d\phi\label{eq:deltayintsb}
\end{align}
and
\begin{align}
F_y(\phi)=&\int_0^\phi f^2_y(\phi')d\phi'\nonumber\\
=&\frac{\phi}{2}\big[(\sqrt{k}l^*)^2+1\big]+\frac{\sin(2\phi)}{4}\big[(\sqrt{k}l^*)^2-1\big]\nonumber\\
&\qquad\qquad\qquad\;\;\;+\sqrt{k}l^*\sin^2\phi.\label{eq:Fy}
\end{align}
Equation~(\ref{eq:deltay}) shows a cubic component and in addition a linear component in $y'^*_0$ whose magnitude depends on the horizontal angle at the IP, showing explicitly the correlation between the two planes when the beam size in the defocusing plane is not negligible.\par
As we are interested in the effect over an ensemble of particles, the expected value of a function $\Psi$ is defined as
\begin{align}
\big\langle \Psi \big\rangle =&\int_{-\infty}^\infty\int_{-\infty}^\infty \Psi\Omega(x'^*_0,y'^*_0) dx'^*_0dy'^*_0,\label{eq:expvaldeltay2}
\end{align}
where $\Omega$ represents a Gaussian distribution in both horizontal and vertical particle angles at the IP.\par
The average particle deviation due to radiation $\big\langle\overline{\Delta y}\big\rangle$ is equal to zero, but in order to explore the correlation of the radiation with $y'^*_0$ the two components, $a$ and $b(x'^*_0)^2$, are evaluated further in this section using $\langle(x'^*_0)^2\rangle=\sigma^2_{x'^*_0}=\epsilon_{Nx}/(\gamma\beta^*_x)$ and later compared with particle tracking results of CLIC~3~TeV~QD0.\par
$\big\langle \big(\overline{\Delta y}\big)^2\big \rangle$ is given by
\begin{align}
\big\langle \big(\overline{\Delta y}\big)^2\big \rangle=&15a^2\bigg(\frac{\epsilon_y}{\beta^*_y}\bigg)^3+3b^2\bigg(\frac{\epsilon_x}{\beta^*_x}\bigg)^2\frac{\epsilon_y}{\beta^*_y}+6ab\bigg(\frac{\epsilon_y}{\beta^*_y}\bigg)^2\frac{\epsilon_x}{\beta^*_x}
\end{align}
and used in Section~\ref{s:mitignonlin} to calculate the maximum theoretical mitigation of the radiation effect and to evaluate the mitigation method.\par
In a similar way, the square of the vertical displacement in Eq.~(\ref{eq:deltays})$, \big(\Delta y\big)^2$, is calculated combining Eqs.~(\ref{eq:averu2}),~(\ref{eq:bendingr}),~(\ref{eq:greenx}) and (\ref{eq:greeny}) by averaging over photons second moment of the energy loss, resulting in Eq.~(\ref{eq:deltay2}).\par
\begin{widetext}
\begin{equation}
\overline{(\Delta y)^2}=\frac{55}{24\sqrt{3}}r_e\frac{\lambda_e}{2\pi}\gamma^5(y'^*_0)^2\int_0^{\sqrt{k}L}\bigg(\big[y'^*_0f_y(\phi)\big]^2+\big[x'^*_0f_x(\phi)\big]^2\bigg)^{3/2}F^2_y(\phi)d\phi.\label{eq:deltay2}
\end{equation}
\end{widetext}
$\big\langle\overline{ (\Delta y)^2}\big\rangle=\sigma^2_{\text{oide}}$ when $\sigma_{x'^*_0}=0$. However, when $\sigma_{x'^*_0}$ is not zero the expected value cannot be calculated analytically, it must be evaluated numerically and it is referred to as $\sigma^2_{\text{2D-oide}}$ below. This is also compared with particle tracking simulations in this section for the case of CLIC~3~TeV.
\subsection{Simulation}
\begin{figure}[h]
 \includegraphics[width=0.33\textwidth,angle=-90]{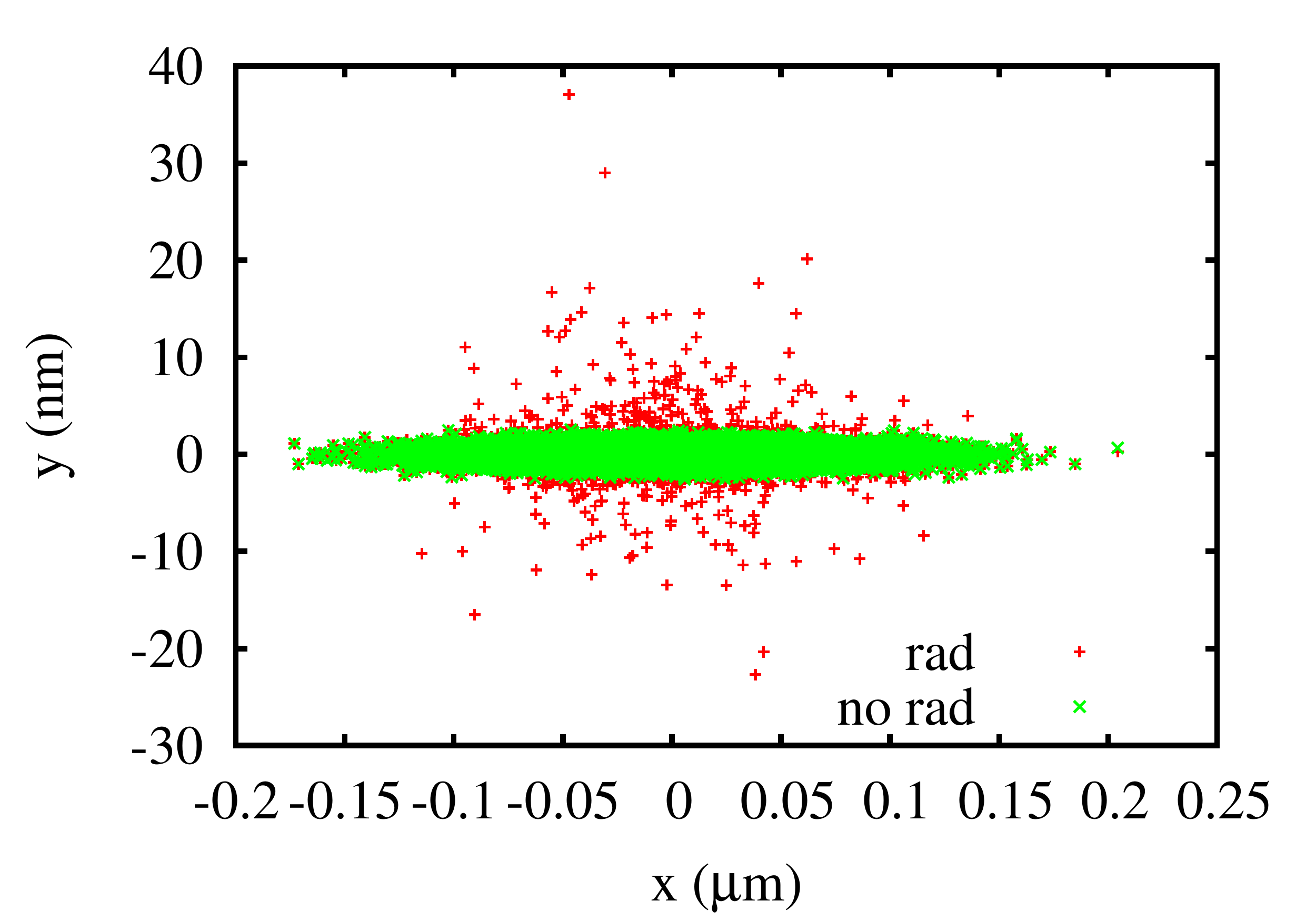}\caption{CLIC~3~TeV beam at the IP after tracking through 	QD0 with and without radiation.}\label{f:CLIC3TeVbeamsizeIP}
\end{figure}

Particle tracking from the entry of QD0 to the IP for CLIC~3~TeV with and without radiation, using PLACET~\cite{Placet}, allows to compute the effects of radiation on the six-dimentional phase space. Figure~\ref{f:CLIC3TeVbeamsizeIP} shows the particle transverse distribution at the IP with and without radiation.\par
\begin{figure}[h]
\includegraphics[width=0.33\textwidth,angle=-90]{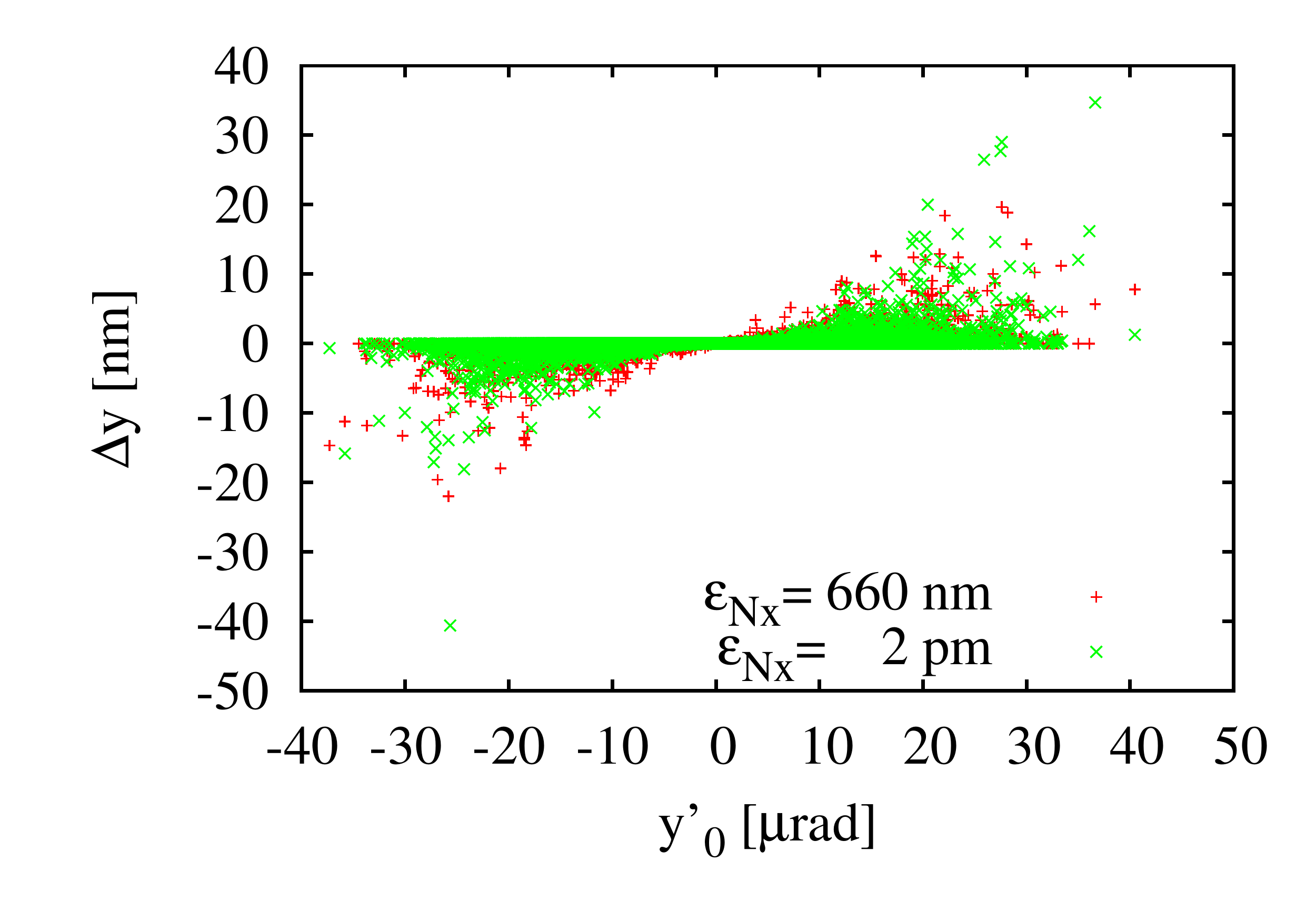}\caption{Correlation between the phase space coordinates $\Delta y,y'$ for CLIC 3 TeV from particle tracking with two different horizontal emittances.}\label{f:correlation}
\end{figure}
Figure~\ref{f:correlation} shows tracking results of $\Delta y = y_{rad}-y_{no\,rad}$ versus the particle angle $y'_0$ at the IP for two different horizontal emittances. It has been included to ilustrate the radiation effect for the nominal normalized horizontal beam emittance, $\epsilon_{Nx}~=~660$~nm, and the case when the horizontal beam size $\sigma_x$ is negligible when compared to the vertical beam size which is achieved by reducing the horizontal emittance, $\epsilon_{Nx}=10^{-4}\epsilon_{Ny}$, i.e. $\epsilon_{Nx}=2$ pm. The horizontal beta twiss parameter at the IP is the same in both trackings, $\beta^*_x~=~6.9$~mm.\par
Fitting results of $\Delta y$ for the two previously mentioned cases are compared  in Table~\ref{t:fitting2} with the analytical evaluation of Eqs.~(\ref{eq:deltay}),~(\ref{eq:deltayintsa}),~(\ref{eq:deltayintsb}) and~(\ref{eq:Fy}).\par
\newcommand\Tstrut{\rule{0pt}{2.8ex}}
{\renewcommand{\arraystretch}{1.2}
\begin{table}[ht]
\begin{tabular}{llcc}\hline\hline
\multicolumn{2}{l}{$\overline{\Delta y}$}& $a$ & $b\,\epsilon_{Nx}/(\gamma\beta_x)$\Tstrut\\[1pt]
&&$[10^{-11}~$m] & $[10^{-11}~$m]\\\hline
$\epsilon_{Nx}= 2$~pm&Theory & $9.0\;\;\;\;\;\;\;\;\;$ & $0\;\;\;\;\;\;\;\;\;$\\
&Tracking& $9.5\pm0.1$ &$-1.3\pm0.3$\\\\
$\epsilon_{Nx}=660$~nm&Theory & $9.0\;\;\;\;\;\;\;\;\;$ & $6.3\;\;\;\;\;\;\;\;\;$\\
&Tracking& $8.5\pm0.1$ &$5.4\pm0.3$\\\hline\hline
\end{tabular}\caption{Coeficients from fitting tracking results and theory Eq.~(\ref{eq:deltay}): $y'_0$ in $[10^{-5}~$rad]~units.}\label{t:fitting2}
\end{table}
}
In the case of $\epsilon_{Nx}=2~$pm, i.e. a negligible effect of the defocusing plane, the result of $a$ differs by 6\% between theory and tracking results, while the result of $b\,\epsilon_{Nx}/(\gamma\beta_x)$ shows an unexpected component from tracking. In the case of $\epsilon_{Nx}=660~$nm, i.e. when the defocusing plane is not negligible, the result of $a$ differs by -6\% between the theory and tracking results, while the result of $b\,\epsilon_{Nx}/(\gamma\beta_x)$ shows -14\% difference. In addition, there is an unexpected 11\% variation in the fitting results of the cubic component $a$ caused by the change of the horizontal emittance in tracking.\par
The contribution to the beam size $\big\langle \overline{(\Delta y)^2}\big\rangle$ from the two theoretical expressions ($\sigma_{\text{oide}}$ and $\sigma_{\text{2D-oide}}$), and the particle tracking are compared in Table~\ref{t:oide2D}, where the errors included in the $\sigma_{\text{2D-oide}}$ results come from numerical precision in the integration of Eq.~(\ref{eq:deltay2}).\par
{\renewcommand{\arraystretch}{1.2}
\begin{table}[ht]
\begin{tabular}{llc}\hline\hline
&\multicolumn{1}{l}{$\big\langle \overline{(\Delta y)^2} \big\rangle^{1/2}$} & [nm]\Tstrut\\[1pt]\hline
$\epsilon_{Nx}= 2$~pm&$\sigma_{\text{oide}}$ &$0.87\;\;\;\;\;\;\;\;\;\;\,$\\
&$\sigma_{\text{2D-oide}}$&$0.87\pm0.03$\\
&Tracking& $0.92\;\;\;\;\;\;\;\;\;\;\,$\\\\
$\epsilon_{Nx}=660$~nm&$\sigma_{\text{2D-oide}}$&$1.02\pm0.03$\\
&Tracking& $1.00\;\;\;\;\;\;\;\;\;\;\,$ \\\hline\hline
\end{tabular}
\caption{Contribution to beam size from radiation evaluated by $\sigma_{\text{oide}}$, $\sigma_{\text{2D-oide}}$ and tracking.}\label{t:oide2D}
\end{table}
}
The case of $\epsilon_{Nx}=2$~pm shows that $\sigma_{\text{oide}}$ and $\sigma_{\text{2D-oide}}$ agree within the numerical precision achieved. The result from tracking is 6\% above the theoretical values. On the other hand, the contribution to radiation calculated from $\sigma_{\text{2D-oide}}$ with the nominal horizontal emittance $\epsilon_{Nx}~=~660$~nm agrees well with the tracking result.\par
At the moment the differences between theory and tracking results for these cases have been attributed to limitations in the particle tracking and radiation simulations.\par
Comparing $\sigma_{\text{2D-oide}}$ to $\sigma_{\text{oide}}$ from Table~\ref{t:oide2D} it is possible to conclude  that including the 2$^\text{nd}$ dimension has an important impact of 11\% in the final IP vertical beam size in CLIC~3~TeV because of the additional 17\% beam size contribution from radiation.\par
\subsection{Mitigating the impact on the beam size by a non-linear corrector scheme}\label{s:mitignonlin}
An ideal compensation system would remove the position change due to radiation $\Delta y = y_\text{rad} -y_{no\,rad}$. In this case, it is possible to remove the average effect due to energy loss $\overline{\Delta y}$ because it correlates with the particle angle in the vertical plane $y'^*_0$.\par
For CLIC~3~TeV, $\big\langle\big(\overline{\Delta y}\big)^2\big\rangle^{1/2}=0.40$~nm, composed of  0.11~nm from the linear and 0.34~nm from the cubic components in $y'^*_0$. Using the expression
\begin{equation}
\frac{\sqrt{\sigma^2_{\text{2D-oide}}-\big\langle\big(\overline{\Delta y}\big)^2\big\rangle}}{\sigma_{\text{2D-oide}}}\label{eq:oidepercent}
\end{equation}
a possible ideal 8\% reduction in the contribution to the Oide effect by removing the correlation. If only one of the two components is removed, then removing the linear component corresponds to 1\% reduction, while removing the cubic components is 6\% reduction in the contribution to beam size due to radiation.\par
Several possibilities arise to achieve the mitigation of the beam size growth, in particular the beam size components due to fringe fields~\cite{PhysRevSTAB.17.101002} show similarities with the components in $\overline{\Delta y}$,  pointing to the possibility to tune the lattice elements to compensate the effect of radiation, possibly using various octupoles.\par
\begin{figure}[floatfix]
\includegraphics[width=0.48\textwidth,angle=0]{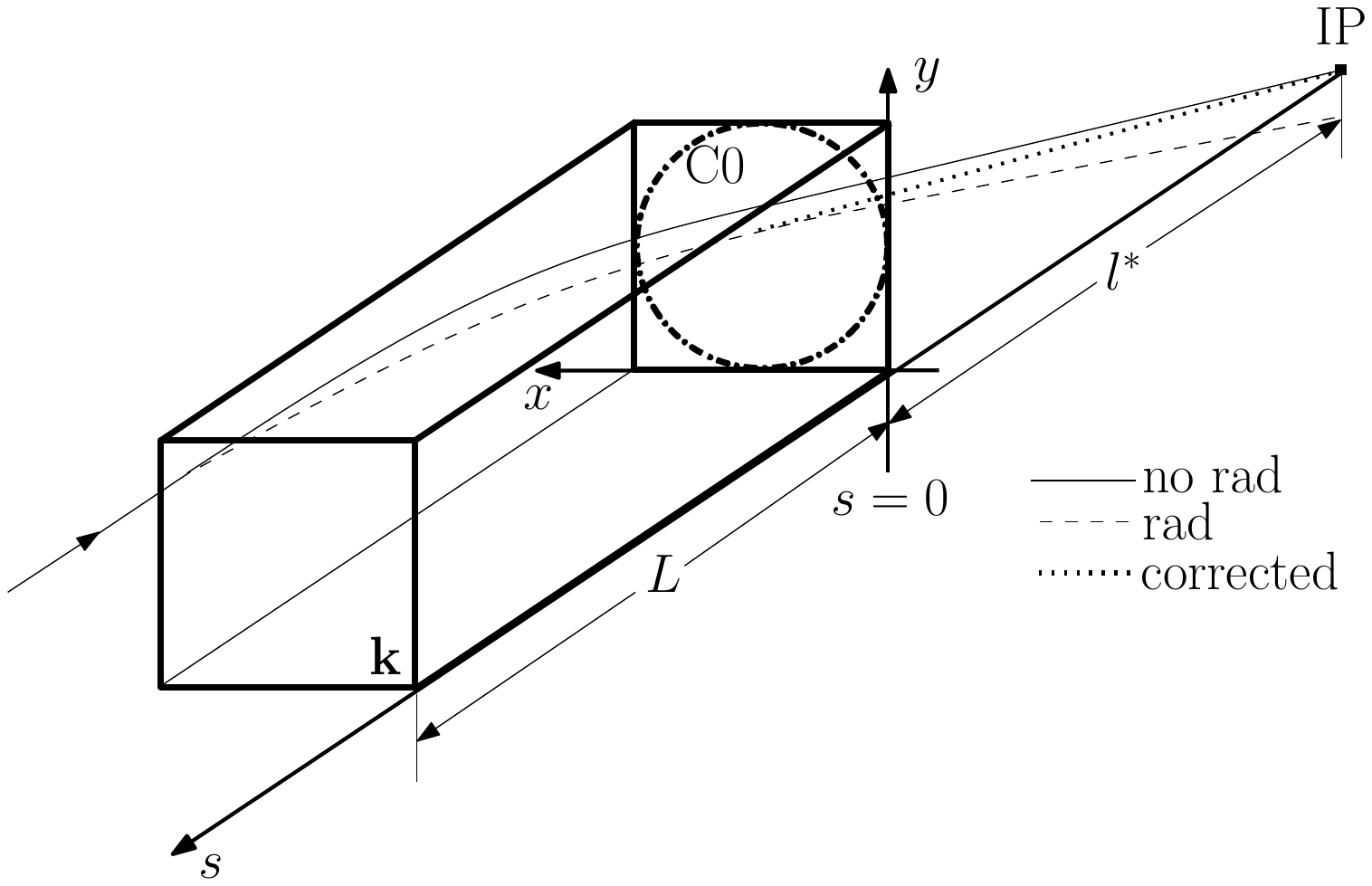}\caption{The solid and dashed lines represent the particle trajectory without and with radiation, respectively. The dotted line represents the correction of the particle trajectory given by C0.}\label{f:corrector}
\end{figure}

In this article, one possible correction scheme is tested consisting in the addition of one octupole magnet C0 in front of the strong focusing quadrupole, placed as in Fig.~\ref{f:corrector}.\par 
The kick given by the corrector, $\Delta y'_{c0}$, changes the vertical position at the IP as $\Delta y_{c0}=L_{c0}\Delta y'_{c0}$, where $L_{c0}$ is the distance from the IP to C0. The displacement induced by the corrector must have opposite direction to the effect of radiation, $\Delta y_{c0}=-\overline{\Delta y}$, therefore
\begin{equation}
\Delta y'_{c0}=-\frac{\overline{\Delta y}}{L_{c0}}.\label{eq:C0deltakick}
\end{equation}
The cubic and linear components of $\overline{\Delta y}$ suggest the use of an octupolar magnet as corrector, therefore, the kick given by an upright octupolar magnet is given by
\begin{flalign}
&x'=-\frac{1}{6}k_3(x^3-3xy^2)\label{eq:kickoctx}\\
&y'=\:\:\:\:\frac{1}{6}k_3(3x^2y-y^3)\label{eq:kickocty}
\end{flalign}
where $k_3$ is the octupole strength. The difference in sign between the linear and cubic components when comparing Eq.~(\ref{eq:kickocty}) with Eq.~(\ref{eq:deltay}) means that only one out of the two could be removed from the beam size by an octupolar corrector, limiting the possible mitigation in this simple approach.\par
\begin{figure}[h]
\includegraphics[width=0.30\textwidth,angle=-90]{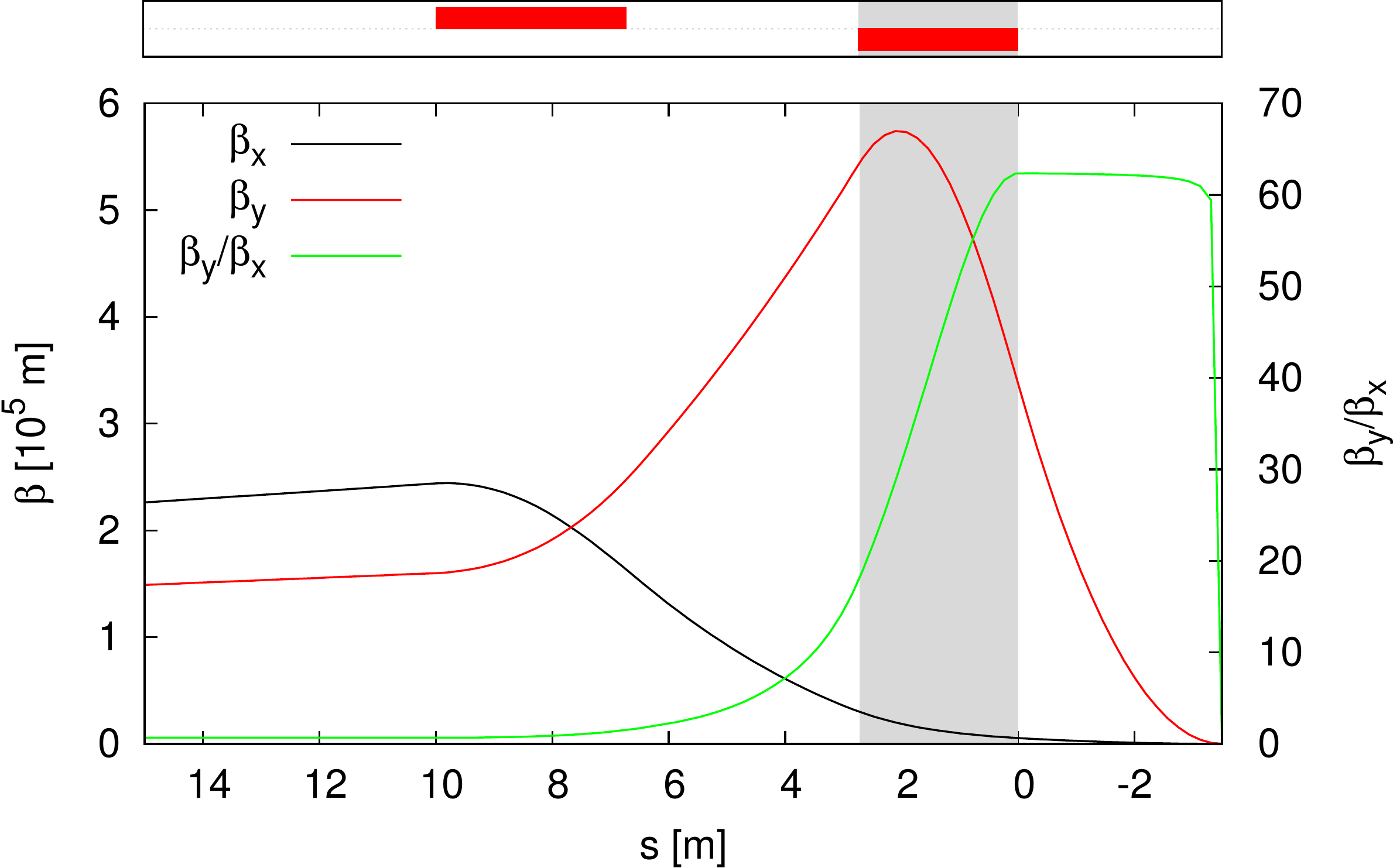}\caption{$\beta$ functions and $\beta_y/\beta_x$ ratio for CLIC 3 TeV FD, QF1 and QD0 in red on top. The dark area is occupied by QD0 and the IP is at $s=-3.5$~m.}\label{f:betaratio}
\end{figure}

The optimum location of the octupolar corrector is in front of the focusing magnet where the vertical beam size is large and the horizontal beam size is small which is achieved when maximizing the $\beta_y/\beta_x$ ratio. Figure~\ref{f:betaratio} shows the horizontal and vertical $\beta$ functions for the CLIC~3~TeV in the FD region, and their ratio, where $\beta_{x}=5.3\times10^{3}$~m and $\beta_{y}=3.3\times10^{5}$~m at C0.\par
Knowing the location of the corrector and matching the kicks in Eqs.~(\ref{eq:C0deltakick}) and (\ref{eq:kickocty}) it is possible to obtain the two values for the octupole strength $k_3$, one to cancel the linear and another to cancel the cubic component,
\begin{align}
\text{Linear:}\quad k_3=&-\frac{2b}{L^4_{c0}}\label{eq:k3lin}\\
\text{Cubic:}\quad k_3=&\:\:\:\:\:\frac{6a}{L^4_{c0}}\label{eq:k3cubic}
\end{align}
where $x'^*_0=x/L_{c0}$ and $y'^*_0=y/L_{c0}$ have been used. For CLIC~3~TeV and $L_{c0}=3.5$~m, the results of Eqs.~(\ref{eq:k3lin}) and~(\ref{eq:k3cubic}) is $k_3~=~-2600$~m$^{-4}$ and $k_3~=~3600$~m$^{-4}$, respectively.\par
In order to try this octupole mitigation in simulations, a CLIC~3~TeV nominal beam with no energy spread is generated at the IP and tracked back to the entrance of QD0 without radiation with the corrector off. This beam is used to study the Oide effect mitigation in QD0 using the previously mentioned scheme by tracking to the IP with radiation. The procedure consists in setting the best octupole strength for a 10~mm length C0.\par
Results from the tracking to the IP with and without radiation with the corrector off, in Table~\ref{t:correctors}, show  that the Oide effect contribution to vertical beam size affects very little the total and peak luminosities, $L_{tot}$ and $L_{peak}$, obtained with Guinea Pig~++~\cite{Schulte:382453}.\par
The best result obtained with the octupole corrector is a vertical beam size reduction by $(-4.3\pm0.5)$\%, equivalent to 6\% reduction in the Oide effect contribution to beam size which agrees with the removal of the cubic component on the beam size.\par
{\renewcommand{\arraystretch}{1.2}
\begin{table}[h]
\begin{tabular}{lccccc}\hline\hline
&$k_3$&$\sigma_x$ & $\sigma_y$ & $L_{tot}$ & $L_{peak}$\\
& [m$^{-4}$] &  [nm] & [nm] & \multicolumn{2}{c}{[$10^{34}$cm$^{-2}\cdot$ s$^{-1}$]}\\\hline
NO RAD & $\;\;\;\;\;\;0$ & 47.45 & 0.69 & 7.7 & 2.9\\
RAD    & $\;\;\;\;\;\;0$ & 47.45 & 1.18 & 7.5 & 2.7 \\
RAD    & 3900 & 47.45 & 1.13 & 7.4 & 2.7 \\\hline\hline
\end{tabular}\caption{Effect of the octupolar corrector on the beam size, total luminosity and peak luminosity.}\label{t:correctors}
\end{table}
}
The C0 strength from tracking is positive also indicating the correction of the cubic component. However it is 8\% bigger than the theoretical value. This has been attributed to limitations in the simulation and the strength optimization.\par
\section{Conclusions}
Synchrotron radiation in the final quadrupole sets a lower limit in the beam size, which is particularly relevant for CLIC~3~TeV. This is normally mitigated by increasing the length of the magnet, nevertheless this produces larger chromatic and geometric aberrations. Previous  results considered only the focal plane of the quadrupole.\par
A new value, $\sigma_{\text{2D-oide}}$, has been derived to include the radiation effect of both the focusing and defocusing planes of the quad showing an increase of 17\% in the contribution to beam size. Theory shows good agreement with simulations using the tracking code PLACET.\par  A closer insight into the beam correlation induced by synchrotron radiation at the IP have revealed a way to mitigate the 2D-Oide effect with octupolar magnets. In general the 2D-Oide average geometric aberrations can be included in codes like MAPCLASS~\cite{githubMapClass2} to mitigate all aberrations simultaneously. The simple case of an octupole is calculated and simulated giving a vertical beam size reduction of $(4.3\pm0.5)$\%, with little or negative impact on luminosity.\par

%

\end{document}